\documentclass[lettersize,journal]{IEEEtran}
\IEEEoverridecommandlockouts
\usepackage{cite}
\usepackage[tbtags]{amsmath}
\usepackage{amssymb,amsfonts}
\usepackage{algorithm}
\usepackage{algorithmic}
\usepackage{graphicx}
\usepackage{textcomp}
\usepackage{xcolor}
\usepackage{caption}
\usepackage{subcaption}
\usepackage{commath}
\usepackage{url} 
\usepackage{multirow}

\DeclareMathOperator*{\argmin}{arg\,min}

\def\BibTeX{{\rm B\kern-.05em{\sc i\kern-.025em b}\kern-.08em
    T\kern-.1667em\lower.7ex\hbox{E}\kern-.125emX}}

\newtheorem{claim}{\textbf{Claim}}

\begin{document}
\title{Cell-Free Beamforming Design for Physical Layer Multigroup Multicasting}

\author{Mahmoud Zaher, Emil Björnson, \IEEEmembership{Fellow, IEEE,} and Marina Petrova, \IEEEmembership{Member, IEEE.}%
\thanks{Mahmoud Zaher, and Emil Björnson are with the Division of Communication Systems, KTH Royal Institute of Technology, 164 40 Stockholm, Sweden (e-mail: mahmoudz@kth.se; emilbjo@kth.se).}

\thanks{Marina Petrova is with the Division of Communication Systems, KTH
Royal Institute of Technology, 164 40 Stockholm, Sweden, and also with the Mobile Communications and Computing Group, RWTH Aachen University, 52062 Aachen, Germany (e-mail: petrovam@kth.se).}
\thanks{This work was supported by the FFL18-0277 grant from the Swedish Foundation for Strategic Research.}}

\maketitle

\begin{abstract}

In many wireless communication applications, it is desirable to transmit the same data to multiple user equipments (UEs). Physical layer multicasting presents an efficient transmission topology to exploit the beamforming capabilities at the transmitting nodes and broadcast nature of the wireless channel to satisfy the demand for the same content from several UEs. An advantage of multicasting is to avoid unnecessary co-channel interference between UEs requesting the same data. The difficulty is to find the suitable beamforming configuration that guarantees an acceptable minimum data rate, among the receiving UE group, to the multicast transmission. This paper addresses the max-min fair multigroup multicast optimization problem and proposes a novel iterative elimination procedure coupled with semidefinite relaxation (SDR) to find the near-optimal rank-1 beamforming vectors in a cell-free massive MIMO (multiple-input multiple-output) network. The proposed optimization procedure significantly improves computational complexity and spectral efficiency compared to common methods that use SDR followed by some randomization procedure and the state-of-the-art difference-of-convex approximation algorithm. The importance of the proposed procedure is that it is applicable to any SDR problem where a low-rank solution is desirable. Further, we propose a low-complexity algorithm that achieves $87\,\%$ of the optimal rank-1 solution at orders-of-magnitude lower computational time.

\end{abstract}

\section{Introduction} \label{sec_intro}
The increase in the demand for wireless services provides a constant drive to push the capacity limits of wireless communication networks. With the growing appetite for multimedia services and the surge in the number of user equipments (UEs), multimedia broadcasting provides an efficient solution to transmit data to a group of UEs. It has received considerable attention since it can support various applications such as television channels' broadcasting over mobile networks, videoconferencing, and emergency broadcast systems. Due to the nature of broadcasting or multicasting, where the UEs are served using the same transmission and time-frequency resources, this technique mitigates the problem of co-channel interference within the group of UEs requesting the same data and provides a remarkable improvement in radio resource utilization \cite{hsu2016joint}. The emergence of massive MIMO (multiple-input multiple-output) as a cornerstone technology to enhance the performance of wireless communication systems offers the opportunity for effective beamforming of a multitude of these multicast signals to different UE groups. Despite the performance gain of massive MIMO, the cell-centric design suffers from the intrinsic inter-cell interference and large pathloss variations of conventional networks \cite{zaher2023soft}. Cell-free massive MIMO represents a post-cellular architecture to provide service for UEs with geographically distributed access points (APs) that cooperate in a joint transmission with no cell boundaries \cite{bjornson2019making,demir2021foundations,zaher2023learning}. It has received significant attention due to its ability to provide uniform coverage within the network area and to suppress inter-cell interference \cite{zaher2023soft}. Such network infrastructure focuses on boosting the poor cell-edge UEs' performance. This is particularly desirable for multicast transmissions as the minimum signal-to-interference-and-noise ratio (SINR) determines the common information rate of a UE group \cite{karipidis2008quality}.

To that end, quality of service (QoS) (guaranteeing a minimum received SNR to every UE) and max-min fair (MMF) (maximizing the smallest received SNR) designs have been first introduced for a single-cell single-multicast group in \cite{sidiropoulos2006transmit}. The idea was later extended in \cite{karipidis2008quality} to consider multiple co-channel multicast groups. The core problem was shown to be NP-hard and approximate solutions were obtained using relaxation techniques on the basis of semidefinite programming (SDP), which is a class of convex optimization problems that can be efficiently solved in polynomial time using interior point methods. The resulting solution matrices found by a generic SDP solver are likely to have a high rank, even if a rank-1 solution exists \cite{ashraphijuo2017multicast}. Hence, algorithmic refinements are needed to identify a practical rank-1 solution. The works in \cite{karipidis2008quality,sidiropoulos2006transmit} suggested different randomization procedures to extract the sub-optimal rank-1 beamforming solutions from the higher-rank solutions attained through SDP. However, as the system size grows large, this method does not scale well in terms of the computational complexity required to test enough random vectors to obtain a satisfactory spectral efficiency (SE) performance. Particularly, in case of having more than one multicast group, a multigroup multicast power control (MMPC) optimization problem needs to be solved for each candidate set of beamforming vectors obtained through this method. Moreover, the approximation quality degrades considerably as the total number of transmit antennas increases \cite{sidiropoulos2006transmit,zhou2017coordinated}.

As opposed to the semidefinite relaxation (SDR) approach and subsequent approximation techniques, a more recent line of research has suggested other iterative optimization procedures to solve the QoS and MMF problems in the context of multigroup multicasting. A low-complexity algorithm based on the Lagrangian dual problem with weak duality is proposed in \cite{dartmann2011low} for the case of multicell multicasting with each cell comprising a single multicast group. It is shown that the proposed solution can achieve satisfactory performance as compared to the SDR with randomization, when considering only the availability of statistical channel state information (CSI) and limited number of UEs per cell. An iterative second-order cone programming (SOCP) procedure is derived in \cite{tran2013conic} for the QoS objective considering a single-cell single-multicast group problem. In \cite{zhou2017coordinated}, a coordinated multigroup multicast MMF problem is formulated and solved using parametric manifold optimization in a multicell network, where each base station (BS) serves a single group of UEs that are associated with its cell. Moreover, the authors in \cite{hsu2016joint} develop three different algorithms to solve the MMF problem with per-cell power constraints while performing joint beamforming among the cooperating BSs. Their numerical evaluation, considering only two cooperating BSs, shows that the difference-of-convex approximation (DCA) algorithm yields the highest SE performance with improved computational complexity compared to the other developed SDR with randomization-based algorithms. The DCA algorithm relies on the successive convex approximation (SCA) technique, which represents the state-of-the-art in multicast beamforming optimization. Other works, such as \cite{wang2016weighted} and references therein, attempt to maximize the weighted sum rate objective in similar network setups, however, we believe that the MMF and QoS utility functions represent more suitable maximization objectives for the multicast problem, as the minimum UE rate in a UE group constitutes the bottleneck for multicast transmissions \cite{park2008capacity}.

Moving towards the massive MIMO multicasting, several works have considered single- and multiple-group multicasting in cooperative and non-cooperative transmission topologies considering large massive arrays equipped at the BSs. In \cite{xiang2014massive}, the asymptotically optimal beamforming vector with non-cooperative BSs is shown to be a linear combination of the channel vectors of the multicast UEs served in its cell, where each cell is considered to serve only a single multicast group. A two-layer beamforming design has been proposed in \cite{sadeghi2017reducing}, where the first layer is tailored to eliminate the multicast inter-group interference while the second layer is designed based on the SCA technique to maximize the SNR of the decoupled multicast groups separately. The work in \cite{dong2020multi,zhang2023ultra} devise low-complexity algorithms, based on the optimal multicast beamforming structure derived in \cite{dong2020multi}, which rely on approximations to the involved parameter matrices of the optimal structure. In \cite{sadeghi2017max}, the maximum ratio (MR) and regularized zero-forcing (RZF) precoding schemes are considered with either the composite channel estimation or the use of a dedicated pilot to each UE.
In addition, \cite{de2022user,li2022spectral} utilize MR and RZF based precoding for the composite multicast channel (sum of the channels in a UE group) with a pilot power allocation scheme that compensates for the pathloss differences between each of the UEs in the group and the serving BS; while \cite{de2022user} considers a single massive MIMO BS and a single multicast channel, whereas \cite{li2022spectral} presents similar results for a relatively small cell-free massive MIMO system with a few APs in close proximity to each other and a small number of UEs. 
The main idea of these papers is to rely on asymptotic expressions to design beamforming vectors that approach the optimum solution when the number of antennas at the BSs grows exceedingly large. Although such design criterion works well in the unicast scheme, the convergence of this approach for the multicast case is slower as the number of antennas increases, especially when the number of UEs per group grows large, i.e., as the multicast problem structure becomes more distant to that of the unicast problem.

In this paper, we tackle the problem of multigroup multicast beamforming optimization. The paper offers an algorithmic contribution and derives a novel iterative optimization procedure that is applicable to a wide range of SDP problems. In particular, we propose a novel optimization procedure to find near-global optimum MMF beamforming vectors in a cell-free network providing service to multiple multicast UE groups. Further, we develop a low-complexity heuristic algorithm that provides an efficient solution to the multigroup multicast MMF problem. We highlight that the proposed procedures are straightforwardly applicable to any multigroup multicast problem and can provide a near-optimal solution, regardless of the network architecture and size. Throughout this paper, we assume perfect CSI acquisition at the transmitting APs to keep the focus on the multicast beamforming optimization problem.
The main contributions of the paper can be summarized as follows:
\begin{itemize}
    \item We propose a new formulation for the QoS problem that directly attempts to minimize the per-AP powers rather than the total power available at all APs, which is common in the previous literature. In this way, we alleviate the need for the redundant total power constraint commonly used in the previous literature, which reduces the number of constraints and the required computational time for the algorithm.

    \item We propose a novel iterative optimization procedure to solve the multigroup multicast MMF problem. The procedure utilizes the SDR technique along with successive elimination of the higher-rank solutions produced by SDR to achieve the near-global optimum rank-1 beamforming solution to the MMF problem as opposed to the state-of-the-art SCA based methods that can only guarantee convergence to a stationary point of the original non-convex problem.

    \item We propose a novel low-complexity heuristic algorithm that provides an effective solution to the multicast beamforming optimization problem at a vastly reduced computational time. The algorithm utilizes a newly proposed phase alignment and UE emphasis procedure that aims at iteratively maximizing the minimum SEs across the multicast UE groups. The proposed heuristic marks the first low-complexity algorithm that is specifically tailored for the multicast problem with computational requirements that are in the milliseconds range.

    \item We provide an extensive performance evaluation of the proposed algorithms and compare them to the state-of-the-art in multigroup multicast beamforming optimization. The numerical results show significant improvements in terms of the minimum achievable SE as well as the computational time required by the algorithms.
\end{itemize}

The rest of the paper is organized as follows: Section~\ref{sys_mod} presents the multigroup multicast communication system model. In Section~\ref{max_min_sec}, the MMF SINR problem is formulated and the relaxed problem is cast as an SDP. The proposed iterative optimization procedure is detailed in Section~\ref{SDR_E}. Section~\ref{heuristic_sec} demonstrates the proposed low-complexity heuristic algorithm. Section~\ref{results} presents numerical results, whereas the main conclusions are summarized in Section~\ref{conc}.

\textbf{Notations:} Lowercase and uppercase boldface letters denote column vectors and matrices, respectively. The symbols $(\cdot)^*$, $(\cdot)^T$, and $(\cdot)^H$ indicate conjugate, transpose and conjugate transpose, respectively. $\mathbb{E}(\cdot)$, tr$(\cdot)$, and $\norm{\cdot}$ denote the expectation, trace, and L$_2$ vector norm, respectively. $\mathbf{I}_M$ represents the $M \times M$ identity matrix. $\mathbf{A} \circ \mathbf{B}$ is the Hadamard product.

\section{System Model} \label{sys_mod}

\begin{figure}
\centering
\setlength{\abovecaptionskip}{0.33cm plus 0pt minus 0pt}
\includegraphics[scale=0.8]{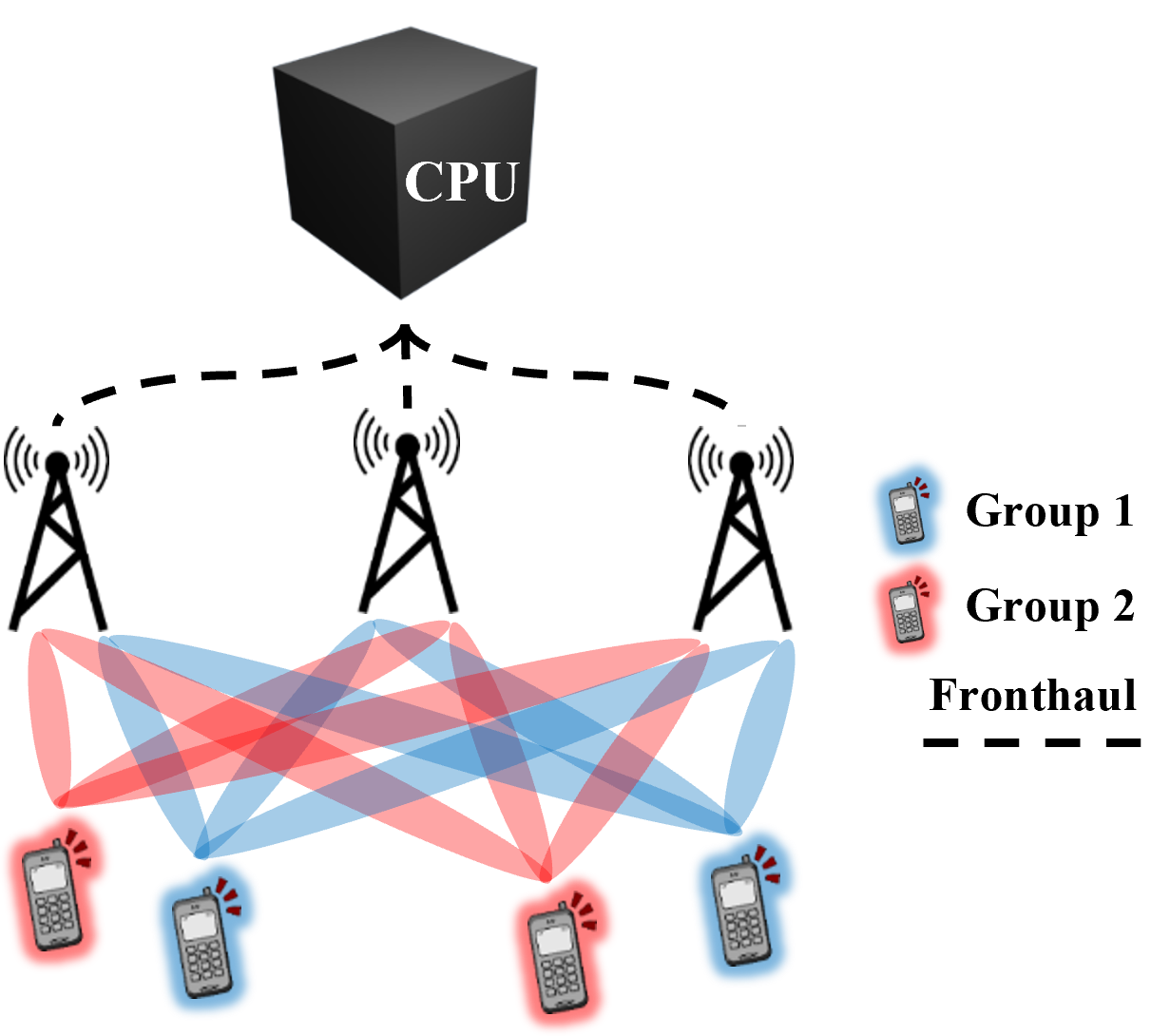}
\caption{An example of a cell-free network with $L = 3$ APs and $G = 2$ multicast groups.}
\label{sys_mod_fig}
\end{figure}

This paper considers a cell-free massive MIMO system with $L$ APs, each equipped with $N$ antennas. The APs jointly serve $K$ single-antenna UEs that are arbitrarily distributed in a large service area. We assume that the UEs are divided into $G$ multicast groups, such that for group $g$, $g = 1, \hdots, G$, all $K_g$ UEs belonging to the group are served with a single multicast transmission. Each UE is assumed to belong to only one multicast group, which entails that $\sum_{j = 1}^GK_j = K$. Note that the adopted system model may be viewed as a generalization to the multi-user downlink unicast problem because it can capture a mix of multicasting and unicasting services, where each unicast UE is modeled as a multicast group having a single UE.
The system model is depicted in Fig.~\ref{sys_mod_fig}.
We consider a standard block fading channel model where the time-varying wideband channels are divided into time-frequency coherence blocks, such that the channels are static and frequency-flat in each block \cite{demir2021foundations}. The channel between UE $k$ in group $g$ and AP $l$ is denoted as $\mathbf{h}_{kg,l} \in \mathbb{C}^{N \times 1}$.

All UEs belonging to the different multicast groups are served by the $L$ APs on the same time-frequency resources. Accordingly, the transmitted signal from AP $l$ is given by
\begin{equation}
    \mathbf{x}_l = \sum_{j = 1}^G\mathbf{w}_{j,l}s_{j},
\end{equation}
where $s_{j}$ denotes the zero-mean unit-variance multicast signal intended for all the UEs of group $j$ and $\mathbf{w}_{j,l}$ represents the corresponding precoding vector. The precoding vectors satisfy short-term power constraints at each AP, which means that the power constraints are satisfied in each coherence block, and not on average. The precoding vectors of AP $l$, $l = 1, \hdots, L$ thus satisfy $\sum_{j = 1}^G\norm{\mathbf{w}_{j,l}}^2 \leq P_{l,\textrm{max}}$, where $P_{l,\textrm{max}}$ represents the maximum transmit power available at AP $l$. The received signal at UE $k$ in group $g$ is computed as
\begin{equation}
\begin{split}
    y_{kg} &= \sum_{l = 1}^L\mathbf{h}_{kg,l}^H\mathbf{x}_l + n_{kg} \\
    &= \sum_{l = 1}^L\mathbf{h}_{kg,l}^H\mathbf{w}_{g,l}s_g + \sum_{l = 1}^L\mathbf{h}_{kg,l}^H\sum_{\substack{j = 1 \\ j \neq g}}^G\mathbf{w}_{j,l}s_{j} + n_{kg},
\end{split}
\end{equation}
where $n_{kg} \sim \mathcal{N}_{\mathbb{C}}(0, \sigma_{kg}^2)$ represents the noise at UE $k$ in group $g$. As a result, an achievable SE of the $k^{\textrm{th}}$ UE of group $g$ under the perfect CSI assumption can be evaluated as
\begin{equation}
    \textrm{SE}_{kg}^{\mathrm{dl}} = \textrm{log}_2\left(1 + \textrm{SINR}_{kg}^{\mathrm{dl}}\right),
\end{equation}
where $\textrm{SINR}_{kg}^{\mathrm{dl}}$ is the effective SINR of UE $k$ in group $g$:
\begin{equation}
    \textrm{SINR}_{kg}^{\mathrm{dl}} = \frac{\left|\sum_{l = 1}^L\mathbf{h}_{kg,l}^H\mathbf{w}_{g,l}\right|^2}{\sum_{\substack{j = 1 \\ j \neq g}}^G\left|\sum_{l = 1}^L\mathbf{h}_{kg,l}^H\mathbf{w}_{j,l}\right|^2 + \sigma_{kg}^2}.
\end{equation}

We define the concatenated channel vector between all APs and UE $k$ in multicast UE group $g$ as $\textbf{h}_{kg} = \left[\textbf{h}_{kg,1}^T, \textbf{h}_{kg,2}^T, \hdots,  \textbf{h}_{kg,L}^T\right]^T \in \mathbb{C}^{LN \times 1}$ and the precoding vector for group $g$ as $\textbf{w}_g = \left[\textbf{w}_{g,1}^T, \textbf{w}_{g,2}^T, \hdots, \textbf{w}_{g,L}^T\right]^T \in \mathbb{C}^{LN \times 1}$. The SINR expression can then be rewritten as
\begin{equation}
\begin{split}
    \textrm{SINR}_{kg}^{\mathrm{dl}} &= \frac{\left|\mathbf{h}_{kg}^H\mathbf{w}_{g}\right|^2}{\sum_{\substack{j = 1 \\ j \neq g}}^G\left|\mathbf{h}_{kg}^H\mathbf{w}_{j}\right|^2 + \sigma_{kg}^2} \\
    &= \frac{\mathbf{h}_{kg}^H\mathbf{w}_{g}\mathbf{w}_{g}^H\mathbf{h}_{kg}}{\sum_{\substack{j = 1 \\ j \neq g}}^G\mathbf{h}_{kg}^H\mathbf{w}_{j}\mathbf{w}_{j}^H\mathbf{h}_{kg} + \sigma_{kg}^2} \\
    &= \frac{\textrm{tr}\left(\mathbf{H}_{kg}\mathbf{W}_{g}\right)}{\sum_{\substack{j = 1 \\ j \neq g}}^G\textrm{tr}\left(\mathbf{H}_{kg}\mathbf{W}_{j}\right) + \sigma_{kg}^2},
\end{split}
\label{SINR_reform}
\end{equation}

\noindent where $\mathbf{H}_{kg} = \mathbf{h}_{kg}\mathbf{h}_{kg}^H$ and $\mathbf{W}_g = \mathbf{w}_g\mathbf{w}_g^H$, $\forall g$. The above SINR formulation makes it suitable to construct the SDP of the multigroup multicast MMF problem.

\section{Max-Min Fair SINR Problem} \label{max_min_sec}

In this section, we formulate the multigroup multicast MMF problem. The goal is to find the set of transmit precoding vectors $\{\mathbf{w}_g$: $\forall g\}$ that maximizes the weighted minimum achievable SINR in \eqref{SINR_reform} for every group in the system under per-AP power constraints. We highlight that the minimum SINR determines what data rate can be utilized for a given multicast group, such that all UEs in the group are able to decode the data. Hence, maximizing the minimum achievable SINR represents an appropriate design criterion for the multicast problem.

\subsection{Problem Formulation}

We define $\mathbf{D}_l = \textrm{blkdiag}\left(\mathbf{Z}_{l1}, \hdots, \mathbf{Z}_{lL}\right)$, with $\mathbf{Z}_{ll} = \mathbf{I}_N$ and $\mathbf{Z}_{li} = \mathbf{0}_{N \times N}, \forall i \neq l$. Hence, the multigroup multicast MMF problem can be written as

\begin{subequations}
\begin{align}
    \mathop{\mathrm{maximize}}\limits_{\{\mathbf{W}_g\}_{g = 1}^G} &\min_{\hspace{0.5em}\substack{\forall k \in \{1, \hdots, K_g\} \\ \forall g \in \{1, \hdots, G\}}}\hspace{0.25em} \frac{1}{\eta_g}\frac{\textrm{tr}\left(\mathbf{H}_{kg}\mathbf{W}_{g}\right)}{\sum_{\substack{j = 1 \\ j \neq g}}^G\textrm{tr}\left(\mathbf{H}_{kg}\mathbf{W}_{j}\right) + \sigma_{kg}^2} \\
    \textrm{s.t.} \quad& \mathbf{W}_g \succeq \mathbf{0} \quad \forall g \in \{1, \hdots, G\}, \\
    \quad& \textrm{rank}\left(\mathbf{W}_g\right) = 1 \quad \forall g \in \{1, \hdots, G\} ,\label{rank1}\\
    \quad& \sum_{g = 1}^G\textrm{tr}\left(\mathbf{D}_l\mathbf{W}_g\right) \leq P_{l,\mathrm{max}} \quad \forall l \in \{1, \hdots, L\},
\end{align}
\label{original_problem}
\end{subequations}

\noindent where $\eta_g$ represents the per-group normalized weight of the SINR target, such that $0 \leq \eta_g \leq 1$, $\forall g$, which allows for having different SINR targets among the different multicast groups. The above problem is non-convex due to the non-convex rank-1 constraint in \eqref{rank1}. A relaxation of the problem is devised by dropping the non-convex rank-1 constraint. Further, the problem is cast in epigraph form by introducing an auxiliary variable $t \geq 0$ to lower bound the worst-case SINR. Accordingly, the relaxed MMF problem is reformulated as
\begin{subequations}
\begin{align}
    &\mathop{\mathrm{maximize}}\limits_{\{\mathbf{W}_g\}_{g = 1}^G, t}\quad t \\
    \textrm{s.t.} \quad &\frac{\textrm{tr}\left(\mathbf{H}_{kg}\mathbf{W}_{g}\right)}{\sum_{\substack{j = 1 \\ j \neq g}}^G\textrm{tr}\left(\mathbf{H}_{kg}\mathbf{W}_{j}\right) + \sigma_{kg}^2} \geq \eta_gt, \\
    &\forall k \in \{1, \hdots, K_g\}, \forall g \in \{1, \hdots, G\},\notag \\
    &\mathbf{W}_g \succeq \mathbf{0} \quad \forall g \in \{1, \hdots, G\}, \\
    &\sum_{g = 1}^G\textrm{tr}\left(\mathbf{D}_l\mathbf{W}_g\right) \leq P_{l,\mathrm{max}} \quad \forall l \in \{1, \hdots, L\}.
\end{align}
\label{max-min_problem}
\end{subequations}

The problem in \eqref{max-min_problem} is quasi-convex, meaning that it cannot be directly solved by general-purpose solvers (since $t$ is a variable). In the next subsection, we present a method to find the optimum solution to the relaxed MMF problem. 

\subsection{Relation-Based Algorithm} \label{relation-based_sec}

Inspired by the work in \cite{karipidis2008quality,hsu2016joint}, we devise the following related QoS problem with a given minimum SINR target to achieve the MMF SINR optimum solution of the relaxed problem. In \cite{karipidis2008quality}, it was assumed that each BS serves only a single multicast group, whereas \cite{hsu2016joint} considers the more general multigroup multicast problem with joint processing among several cooperating BSs. However, in this paper, we directly consider the per-AP power constraints in the optimization objective, which is different from the approach of \cite{hsu2016joint} that utilizes the less stringent total power constraint at all APs, and then includes the per-AP power budget as separate constraints.

The corresponding QoS problem to the original MMF problem \eqref{original_problem} is formulated directly in terms of the normalized per-AP powers as

\begin{subequations}
\begin{align}
    \mathop{\mathrm{minimize}}\limits_{\{\mathbf{W}_g\}_{g = 1}^G} &\max_{\hspace{0.5em}\forall l \in \{1, \hdots, L\}} \frac{1}{P_{l,\mathrm{max}}}\sum_{g = 1}^G\textrm{tr}\left(\mathbf{D}_l\mathbf{W}_g\right) \\
    \textrm{s.t.} \quad &\frac{1}{\eta_g}\frac{\textrm{tr}\left(\mathbf{H}_{kg}\mathbf{W}_{g}\right)}{\sum_{\substack{j = 1 \\ j \neq g}}^G\textrm{tr}\left(\mathbf{H}_{kg}\mathbf{W}_{j}\right) + \sigma_{kg}^2} \geq \gamma, \\
    &\forall k \in \{1, \hdots, K_g\}, \forall g \in \{1, \hdots, G\},\notag \\
    &\mathbf{W}_g \succeq \mathbf{0} \quad \forall g \in \{1, \hdots, G\}, \\
    & \textrm{rank}\left(\mathbf{W}_g\right) = 1 \quad \forall g \in \{1, \hdots, G\},
\end{align}
\label{QoS_problem1}
\end{subequations}

\noindent where $\gamma$ represents the SINR target. As in the case of the MMF problem, we drop the non-convex rank-1 constraint to obtain the relaxed convex QoS problem. Subsequently, through the introduction of the auxiliary variable $x \geq 0$ to upper bound the maximum allowable normalized per-AP power,  the relaxed QoS problem is cast in epigraph form as
\begin{subequations}
\begin{align}
    &\mathop{\mathrm{minimize}}\limits_{\{\mathbf{W}_g\}_{g = 1}^G, x}\quad x \\
    \textrm{s.t.} \quad &\frac{\textrm{tr}\left(\mathbf{H}_{kg}\mathbf{W}_{g}\right)}{\sum_{\substack{j = 1 \\ j \neq g}}^G\textrm{tr}\left(\mathbf{H}_{kg}\mathbf{W}_{j}\right) + \sigma_{kg}^2} \geq \eta_g\gamma, \\
    &\forall k \in \{1, \hdots, K_g\}, \forall g \in \{1, \hdots, G\}, \notag \\
    &\mathbf{W}_g \succeq \mathbf{0} \quad \forall g \in \{1, \hdots, G\}, \\
    &\sum_{g = 1}^G\textrm{tr}\left(\mathbf{D}_l\mathbf{W}_g\right) \leq P_{l,\mathrm{max}}x \quad \forall l \in \{1, \hdots, L\}.
\end{align}
\label{QoS_problem2}
\end{subequations}

Problem \eqref{QoS_problem2} is a standard SDP problem that can be solved with an optimization tool, such as CVX \cite{cvx}. That being said, we utilize the relation between the relaxed QoS and MMF problems to solve the relaxed MMF problem. The idea is based on the fact that the optimal objective values of the relaxed MMF and QoS problems are monotonically non-decreasing in the per-AP power budgets and the SINR target, respectively \cite{karipidis2008quality,hsu2016joint}. For that reason, a bisection search is performed over that SINR target such that the resulting required per-AP power is equal to the available power budget at the APs. Let the relaxed MMF problem in \eqref{max-min_problem} be denoted as $\mathcal{F}$ with an optimal objective value of $t^\star$, and the relaxed QoS problem in \eqref{QoS_problem2} as $\mathcal{Q}$. Further, we define $\mathbf{p} = [P_{1,\mathrm{max}}, \hdots, P_{L,\mathrm{max}}]^T \in \mathbb{R}^L$ as the vector of per-AP transmit power budgets, and $\mathbf{g} = [\eta_1, \hdots, \eta_G] \in \mathbb{R}^G$ as the normalized weight vector for the per-group SINR target. The relation between the relaxed QoS and MMF problems is provided in the following claim.
\begin{claim}
    \textit{Problems} $\mathcal{Q}$ \textit{and} $\mathcal{F}$ \textit{are inverse related as}
    \begin{align}
        &1 = \mathcal{Q}\left(\mathcal{F}\left(\mathbf{g}, \mathbf{p}\right)\cdot\mathbf{g},\mathbf{p}\right), \\
        &t^\star = \mathcal{F}\left(\mathbf{g},\mathcal{Q}\left(t^\star\mathbf{g}, \mathbf{p}\right)\cdot\mathbf{p}\right).
    \end{align}
\end{claim}

The logic here is that if the minimum power that can be achieved for a given QoS problem is equal to the power budget at the APs, then the target SINR of the QoS problem is the MMF SINR that can be achieved for that power budget. The above problem is likely to result in optimal solution matrices $\{\mathbf{W}_g\}_{g = 1}^G$ that do not satisfy the rank-1 constraint of the original problem \eqref{original_problem}. In fact, standard interior-point algorithms for solving SDPs will always return the solution with maximum rank among all optimum solutions \cite{luo2010semidefinite}. The reason is that low-rank matrix solutions to SDP problems belong to the boundary of the feasible set and not the interior. Note that a rank-1 beamforming solution where the cooperating APs transmit a single common independent signal to each group utilizing spatial multiplexing is deemed practical due to ease of implementation as compared to a higher-rank transmission.

To solve this issue, previous works have generally used a randomization procedure \cite{karipidis2008quality} to extract a rank-1 beamforming solution. Particularly for the multigroup multicast scenario, an MMPC problem needs to be solved for each set of candidate beamforming vectors produced using randomization in order to yield the optimum power scaling factors that result in the max-min SINR for that set, while still satisfying the per-AP power constraints. In addition, an increasingly large number of candidate sets needs to be tested when increasing the network size and number of connected UEs for a satisfactory SE performance, which results in high computational requirements.

\section{Proposed Successive Elimination Algorithm} \label{SDR_E}

As discussed in Section \ref{relation-based_sec}, when applying SDR to the multigroup multicast MMF problem, the resulting solution matrices will most likely have high rank. Moreover, the randomization procedures in the literature to find an approximate rank-1 solution are particularly unsuitable in the case of multigroup multicasting and the MMPC problem which needs to be solved for each randomization sample scales approximately quadratically \cite{boyd2004convex} with the total number of transmit antennas and UEs. This necessitates the development of an efficient procedure to extract a near-optimal rank-1 solution from the higher-rank solution obtained via SDR. To that end, we propose a novel \emph{Successive Elimination Algorithm}, that is able to find the near-optimal rank-1 solution to the multigroup multicast MMF problem. 

\subsection{Penalized QoS Problem Formulation}

The reformulated multigroup multicast QoS problem in \eqref{QoS_problem2}, along with the bisection search over the different SINR targets, represent the first step of the proposed algorithm. In this way, we are able to achieve a higher-rank solution for the MMF problem by solving a sequence of QoS problems with different SINR targets, until convergence to the SINR target that results in satisfying the power budget constraints at all APs with equality. Afterwards, an iterative procedure to eliminate the higher-rank solutions is performed. The elimination is done through the successive penalization of the eigenvector directions corresponding to the second-largest eigenvalues of the higher-rank solution matrices for each of the multicast UE groups. At each iteration, the following optimization problem is solved:
\begin{subequations}
\begin{align}
    &\mathop{\mathrm{minimize}}\limits_{\{\mathbf{W}_g\}_{g = 1}^G, x}\quad x \\
    \textrm{s.t.} \quad &\frac{\textrm{tr}\left(\mathbf{H}_{kg}\mathbf{W}_{g}\right)}{\sum_{\substack{j = 1 \\ j \neq g}}^G\textrm{tr}\left(\mathbf{H}_{kg}\mathbf{W}_{j}\right) + \sigma_{kg}^2} \geq \eta_g\gamma, \\
    &\forall k \in \{1, \hdots, K_g\}, \forall g \in \{1, \hdots, G\},\notag \\
    &\mathbf{W}_g \succeq \mathbf{0} \quad \forall g \in \{1, \hdots, G\}, \\
    &\sum_{g = 1}^G\left(\textrm{tr}\left(\mathbf{D}_l\mathbf{W}_g\right) + \zeta\sum_{i_g} \left(\mathbf{u}_{i_g}^g\right)^H\mathbf{W}_g\mathbf{u}_{i_g}^g\right) \leq P_{l,\mathrm{max}}x \label{penalty}\\
    &\forall l \in \{1, \hdots, L\}, \notag
\end{align}
\label{QoS_elimination}
\end{subequations}

\noindent where $\zeta$ represents the penalty factor and the $\mathbf{u}_{i_g}^g$, $\forall i_g$, $\forall g$, represent the eigenvectors corresponding to the second-largest eigenvalues that are determined from the higher-rank solutions $\{\mathbf{W}_g\}_{g = 1}^G$ of the previous iterations. 

\subsection{Successive Elimination Algorithm} \label{sec_SEA}

In general, a lower-rank solution to an SDP problem lies on the boundary of the feasible set. As a result, any standard SDP solver will always favor a higher-rank solution over its lower-rank counterpart, even if both can achieve the same objective value. Since we are interested in achieving the optimal rank-1 beamforming solution, we develop an iterative procedure to eliminate all higher-rank solutions that result in an optimal objective value that is greater than or equal to that achieved with the optimum rank-1 solution. The procedure is summarized in Algorithm \ref{alg1}. Note that $P_T$ represents the sum of all AP power budgets; that is $P_T = \sum_{l = 1}^LP_{l,\mathrm{max}}$, $\mathcal{B}(1)$ denotes the first member in the set $\mathcal{B}$, and $0 < \kappa < 1$ determines the elimination interval width.

\begin{algorithm}
        \caption{Successive Elimination Algorithm}
        \label{alg1}
        \noindent\textbf{Input:} Specify the interval $\left[\gamma_\mathrm{lo}, \gamma_\mathrm{up}\right]$ containing the optimum objective value of \eqref{max-min_problem}, $\gamma_\mathrm{lo} = 0$ and $\gamma_\mathrm{up} = \min\limits_{k, g} \frac{P_T\norm{\mathbf{h}_{kg}}^2}{\sigma_{kg}^2}$. Set the solution tolerance to $\epsilon>0$.
	\begin{algorithmic}[1]
            \REPEAT \label{part1}
            \STATE Set $\gamma \leftarrow \left(\gamma_\mathrm{lo} + \gamma_\mathrm{up}\right) / 2$.
		\STATE Solve problem \eqref{QoS_problem2} using the value of $\gamma$.
		\IF{$x > 1$ \OR problem \eqref{QoS_problem2} is infeasible}
                \STATE Set $\gamma_\mathrm{up} \leftarrow \gamma$.
            \ELSE
            \STATE Set $\gamma_\mathrm{lo} \leftarrow \gamma$.
            \ENDIF
            \UNTIL $\gamma_\mathrm{up} - \gamma_\mathrm{lo} < \epsilon$
            \STATE Set $\mathcal{B} \leftarrow \{\mathbf{W}_g~|~\text{rank}\left(\mathbf{W}_g\right) \neq 1\}$.
            \STATE Set $i_g \leftarrow 1, \forall g$.
            \WHILE{$\mathcal{B} \neq \emptyset$}
            \STATE Set $\mathbf{u}_{i_g}^g \leftarrow$ Eigenvector corresponding to second-largest eigenvalue of $\mathcal{B}\left(1\right)$, where $g$ is the group index of $\mathcal{B}\left(1\right)$.
            \STATE Set $\gamma_\mathrm{lo} \leftarrow \min\left(\kappa\gamma, \gamma - 1\right)$. \label{kappa_interval}
            \REPEAT
            \STATE Set $\gamma \leftarrow \left(\gamma_\mathrm{lo} + \gamma_\mathrm{up}\right) / 2$.
            \STATE Solve problem \eqref{QoS_elimination} using the value of $\gamma$.
            \IF{$x > 1$ \OR problem \eqref{QoS_elimination} is infeasible}
                \STATE Set $\gamma_\mathrm{up} \leftarrow \gamma$.
            \ELSE
            \STATE Set $\gamma_\mathrm{lo} \leftarrow \gamma$.
            \ENDIF
            \UNTIL $\gamma_\mathrm{up} - \gamma_\mathrm{lo} < \epsilon$
            \STATE Set $\mathcal{B} \leftarrow \{\mathbf{W}_g~|~\text{rank}\left(\mathbf{W}_g\right) \neq 1\}$.
            \STATE Set $i_g \leftarrow i_g + 1$.
            \ENDWHILE
\end{algorithmic}
\textbf{Output:} The near-optimal rank-$1$ solution matrices.
\end{algorithm}

Algorithm \ref{alg1} starts by specifying the search interval that contains the optimal objective value of the relaxed MMF problem in \eqref{max-min_problem}. Subsequently, a bisection search is performed, where at each iteration, the relaxed QoS problem in \eqref{QoS_problem2} is solved at the midpoint of the search interval. If the solution results in an objective value $x$ (normalized upper bound for the AP powers) that is greater than $1$, or if the problem is infeasible, the upper half of the search interval is discarded; otherwise, the lower half is discarded. These steps are repeated until convergence; that is $\gamma_\mathrm{up} - \gamma_\mathrm{lo} < \epsilon$. Now we arrive at a solution to the relaxed MMF problem where the solution matrices have a rank that is most likely higher than $1$. Next, the elimination procedure of the higher-rank solutions is done such that at each iteration, a higher-rank solution matrix to a given multicast group is penalized in all subsequent iterations. The penalty is in the form of the quadratic matrix product in \eqref{penalty}, alongside the penalty factor $\zeta$, which ensures the elimination of that solution, i.e., generates enough penalty to push the solver away from that solution. The procedure is repeated until all higher-rank solutions that are produced as output by the SDP are eliminated, yielding the desired near-optimal rank-1 solution to the MMF problem.

We highlight two fundamental advantages of the proposed algorithm compared to previous methods. The first advantage is that it does not require an initialization with a feasible solution, which is necessary for algorithms that rely on the SCA technique (i.e., the state-of-the-art in multicast beamforming optimization). Specifically, when the number of UEs grows large, it might be hard to find an effective low-complexity solution to use as initialization. The second advantage is that the proposed elimination procedure to extract the rank-1 solution can perform the bisection search over possible SINR targets within a relatively small interval. This allows for a significant reduction in the computational time of the elimination procedure iterations as compared to the first solution (the conventional higher-rank SDP solution). The reason is that the upper bound of the interval will be the optimum $\gamma$ achieved in the previous iteration, whereas the lower bound can be chosen only a small separation away from it (Step \ref{kappa_interval}). In Section \ref{complexity}, we will demonstrate the computational requirements of the proposed algorithm and compare it to the state-of-the-art.

\subsection{Near-Global Optimality}

Since the multicast MMF and QoS are non-convex and NP-hard problems, it is not possible to obtain their global optimal solution efficiently. The state-of-the-art SCA technique starts from an initial feasible solution and can only achieve a local optimal solution which might have a large gap to the global optimum. In contrast, the proposed SEA starts from an infeasible (higher-rank) solution and approaches the feasible set by introducing the penalty terms in \eqref{penalty}. As a result, the obtained solution will be on the boundary of the feasible set where the global optimal rank-1 solution to the original problem lies, as previously stated. In fact, the obtained rank-1 solution from the SEA represents the global optimum to the penalized QoS problem, where the optimality gap due to relaxation becomes $0$. By utilizing $\mathbf{W}_g = \mathbf{w}_g\mathbf{w}_g^H$, $\forall g$ and reformulating the penalty term accordingly, the obtained solution is the global optimum to the non-convex penalized QoS problem, which can be reformulated as:
\begin{subequations}
\begin{align}
    \mathop{\mathrm{minimize}}\limits_{\{\mathbf{w}_{g,l}\}_{\forall g, l}} &\hspace{-0.5em}\max_{\hspace{0.5em}\forall l \in \{1, \hdots, L\}} \frac{1}{P_{l,\mathrm{max}}}\sum_{g = 1}^G\Biggl(\norm{\mathbf{w}_{g,l}}^2 + \zeta\sum_{i_g}|\mathbf{u}_{i_g}^H\mathbf{w}_{g}|^2\Biggr) \\
    \textrm{s.t.} &\quad \frac{1}{\eta_g}\frac{\left|\mathbf{h}_{kg}^H\mathbf{w}_{g}\right|^2}{\sum_{\substack{j = 1 \\ j \neq g}}^G\left|\mathbf{h}_{kg}^H\mathbf{w}_{j}\right|^2 + \sigma_{kg}^2} \geq \gamma, \\
    &\quad \forall k \in \{1, \hdots, K_g\}, \forall g \in \{1, \hdots, G\}.\notag
\end{align}%
\label{QoS_problem1}
\end{subequations}%

The $\mathbf{u}_{i_g}$, $\forall i_g, \forall g$ in the additive penalty terms correspond to the eigenvectors associated with the second-largest eigenvalues of higher-rank solutions obtained in the previous iterations. Since the solution matrices to the relaxed SDP are Hermitian positive semidefinite with distinct non-zero eigenvalues, the eigenvectors of each matrix are orthogonal to each other, minimizing the impact on the achieved objective value of the original QoS problem. Note that the second-largest eigenvectors cannot yield an optimal rank-1 solution, as the dominant eigenvector of that solution will always provide a higher objective value. As a result, penalizing these eigenvectors has little effect on later iterations, ultimately leading to a near-global optimal rank-1 solution. In Section \ref{perf_analysis}, we will analyze the effect of these penalty terms on the optimality gap numerically.

\section{Phase Alignment Heuristic Algorithm} \label{heuristic_sec}

In this section, we propose a novel low-complexity heuristic algorithm that effectively decides on the multigroup multicast beamforming vectors at orders-of-magnitude lower computational cost compared to the existing as well as proposed optimization solutions. As with unicast transmission, multicast beamforming has to take into account the following three components: $1)$ Interference mitigation, $2)$ Boosting the desired signal, and $3)$ Power allocation. A primary difference is that each precoder is responsible for boosting the desired signal gain for the entire multicast group, which requires additional ministration in the precoder design. To simplify the notation, hereafter we assume the same per-AP maximum transmit powers, $P_{l,\textrm{max}} = P_{\textrm{max}}$, $\forall l$, and the same noise variance at all the UEs, $\sigma_{kg} = \sigma$, $\forall k, g$.

To incorporate these three components in our precoder design, the multigroup multicast precoder of group $g$ can be broadly expressed as
\begin{equation}
    \mathbf{w}_g = \sqrt{\alpha_g}\frac{\bar{\mathbf{w}}_g}{\norm{\bar{\mathbf{w}}_g}},
\end{equation}
where $\alpha_g$ is the total power scaling factor of group $g$ and $\bar{\mathbf{w}}_g$ is given by
\begin{equation}
    \bar{\mathbf{w}}_g = \mathbf{R}_{\mathrm{int}, g}^{-1}\mathbf{d}_g,
\end{equation}
where $\mathbf{R}_{\mathrm{int}, g}$ is the inter-group interference plus noise matrix associated with group $g$ and $\mathbf{d}_g$ is the vector that steers the precoder so as to boost the desired signal of the group, and will be detailed later. With regard to interference mitigation, an advantage of multicasting over unicasting is that it avoids unnecessary co-channel interference within the group of UEs requesting the same data content. Accordingly, following an RZF-like structure, the matrix $\mathbf{R}_{\mathrm{int}, g}$ is defined as
\begin{equation}
    \mathbf{R}_{\mathrm{int}, g} = \left(\sum\limits_{\substack{j = 1 \\ j \neq g}}^G\mathbf{H}_{j}\mathbf{H}_{j}^H + \sigma^2\mathbf{I}_{LN}\right),
\end{equation}
where $\mathbf{H}_j = [\mathbf{h}_{1j}, \hdots, \mathbf{h}_{K_jj}] \in \mathbb{C}^{LN \times K_j}$ is the group channel matrix of group $j$. The effective group channel matrix $\bar{\mathbf{H}}_g = [\bar{\mathbf{h}}_{1g}, \hdots, \bar{\mathbf{h}}_{K_gg}]$ of group $g$, which represents the group channel directions after interference cancellation, is thus computed as
\begin{equation}
    \bar{\mathbf{H}}_g = \mathbf{R}_{\mathrm{int}, g}^{-1}\mathbf{H}_g.
\end{equation}

As for the second component of our heuristic multicast precoder, each beamforming vector has to encompass boosting the desired signal of an entire multicast UE group. As such, the vector $\mathbf{d}_g$ is designed as a weighted sum of the columns of the effective group channel matrix $\bar{\mathbf{H}}_g$. As stated earlier, this component of the beamforming vectors represents the primary difference to the unicast scenario and requires additional consideration in the precoder design. For that purpose, we develop an iterative phase alignment and UE emphasis procedure that is tailored to boost the desired signal of the UE with the minimum SINR in a given multicast group in every iteration. The full multigroup multicast heuristic algorithm is detailed in Algorithm \ref{alg2}.
Note that $\boldsymbol{\rho}_{\mathrm{rt},g} = [\sqrt{\rho_{1g}^\star}, \hdots, \sqrt{\rho_{K_gg}^\star}]^T$, where $\rho_{kg}^\star$ denotes the optimum max-min unicast power for UE $k$ in group $g$. The matrix $\mathbf{P}_g \in \mathbb{C}^{LN \times K_g}$ is the row-wise replication of $\boldsymbol{\rho}_{\mathrm{rt},g}^T$, that is $\mathbf{P}_g = [\boldsymbol{\rho}_{\mathrm{rt},g}, \hdots, \boldsymbol{\rho}_{\mathrm{rt},g}]^T$. $S$ is the number of iterations for the phase alignment and UE emphasis scaling procedure, and $r \geq 1$ is the constant UE emphasis scaling factor. The matrix $\widetilde{\mathbf{H}}_g \in \mathbb{C}^{LN \times K_g}$ denotes the normalized effective group channel matrix such that column $k$, denoted as $\tilde{\mathbf{h}}_{kg}$, $\forall k \in \{1, \hdots, K_g\}$ is given by $\tilde{\mathbf{h}}_{kg} = \bar{\mathbf{h}}_{kg}/\norm{\bar{\mathbf{h}}_{kg}}$. $\mathbf{x}^{(y)}$ and $\mathbf{X}^{(y_1,y_2)}$ denote the $y^{th}$ element in vector $\mathbf{x}$ and element $(y_1,y_2)$ in matrix $\mathbf{X}$, respectively. $\angle(\cdot)$ denotes the phase angle.

Since we adopt a cell-free network setup, the difference in the average channel gains of the serving APs to a given UE makes the design of a heuristic that aims to approach the max-min SINR objective particularly challenging. On the one hand, each AP should try to equalize the desired signal gains of the UE groups. On the other hand, further away UEs should not be given more power by an AP, but rather be focused on by other closer APs. To alleviate this problem, the optimum max-min unicast powers are used to initialize the effective group channel matrix weights (Step \ref{initialization}). For the unicast initialization, we utilize RZF precoding such that the normalized precoding vector for UE $k$ in group $g$ is given by
\begin{equation}
    \mathbf{w}_{kg}^\mathrm{uni} = \frac{\bar{\mathbf{w}}_{kg}^\mathrm{uni}}{\big\lVert\bar{\mathbf{w}}_{kg}^\mathrm{uni}\big\rVert},
\end{equation}
with $\bar{\mathbf{w}}_{kg}^\mathrm{uni}$ computed as
\begin{equation}
    \bar{\mathbf{w}}_{kg}^\mathrm{uni} = \left(\sum\limits_{g = 1}^G\sum\limits_{k = 1}^K\mathbf{h}_{kg}\mathbf{h}_{kg}^H + \frac{\sigma^2}{P_{\textrm{max}}}\mathbf{I}_{LN}\right)^{-1}\mathbf{h}_{kg}.
\end{equation}

Then, the centralized unicast max-min power allocation problem is formulated as
\begin{subequations}
    \begin{align}
        \mathop{\mathrm{maximize}}\limits_{\rho_{kg}, \forall k, g} &\min_{\substack{\forall k \in \{1, \hdots, K_g\} \\ \forall g \in \{1, \hdots, G\}}} \frac{\rho_{kg}|\mathbf{h}_{kg}^H\mathbf{w}_{kg}^{\mathrm{uni}}|^2}{c_{kg} + \sigma^2},\\
        \textrm{s.t.} \quad &\sum\limits_{g = 1}^G\sum\limits_{k = 1}^{K_g}\rho_{kg}\big\lVert\mathbf{w}_{kg, l}^\textrm{uni}\big\rVert^2 \leq P_{\textrm{max}} \quad \forall l \in \{1, \hdots, L\},
    \end{align}
    \label{unicast_max-min}
\end{subequations}
where
\begin{equation}
    c_{kg} = \sum\limits_{\substack{j=1 \\ j \neq g}}^G\sum\limits_{i = 1}^{K_j}\rho_{ij}|\mathbf{h}_{kg}^H\mathbf{w}_{ij}^{\mathrm{uni}}|^2 + \sum\limits_{\substack{i = 1 \\ i \neq k}}^{K_g}\rho_{ig}|\mathbf{h}_{kg}^H\mathbf{w}_{ig}^{\mathrm{uni}}|^2,
\end{equation}
represents the interference experienced by UE $k$ in group $g$ under the unicast transmission scheme and $\mathbf{w}_{kg, l}^\textrm{uni} \in \mathbb{C}^{N \times 1}$ is the portion of the normalized unicast precoding vector corresponding to AP $l$. This problem can be efficiently solved using the fixed-point algorithm in \cite[Alg.~$7.3$]{demir2021foundations} to get the optimum unicast powers $\rho_{kg}^\star$, $\forall k, g$.

Afterwards, we compute the matrix $\mathbf{R}_g$, which characterizes the multiplication of the different effective group channel components with those of the precoder. Next, in Steps \ref{k_min}-\ref{UE_min}, the phase alignment procedure identifies the index and component of the worst UE within the group. The effective channel weight of the minimum UE is then scaled up and phase-tuned such that the product of the effective channel of the minimum UE with its corresponding component in the precoder is phase-aligned to the sum of the product of the effective channel of the minimum UE with all other components in the precoder (Steps \ref{theta}-\ref{rho_update}). The matrix $\mathbf{R}_g$ is then updated and the procedure is repeated for each multicast group for a number of iterations $S$.

Finally, after computing all group precoders, the precoding vectors are equally scaled so as to satisfy the most violated per-AP power. We have observed numerically that it is the per-AP power allocation to the different groups, computed as the squared norm of the portion of the precoding vector corresponding to a given AP $l$: $\alpha_g\norm{\mathbf{w}_{g,l}}^2$ for group $g$, is what makes a significant difference in the achievable max-min SE, and not the total power. As a result, we perform equal total power allocation, which entails $\alpha_g = \alpha_G$, $\forall g$.

\begin{algorithm}[t!]
        \caption{Phase Alignment Heuristic Algorithm}
        \label{alg2}
        \noindent\textbf{Input:} The channel between UE $k$ in group $g$ and AP $l$ $\mathbf{h}_{kg,l},~\forall k,~\forall g,~\forall l$. The per-AP power budget $P_{l,\mathrm{max}},~\forall l$.
	\begin{algorithmic}[1]
            \STATE Compute $\bar{\mathbf{w}}_{kg}^\mathrm{uni} = \left(\sum\limits_{g = 1}^G\sum\limits_{k = 1}^K\mathbf{h}_{kg}\mathbf{h}_{kg}^H + \frac{\sigma^2}{P_{\textrm{max}}}\mathbf{I}_{LN}\right)^{-1}\mathbf{h}_{kg}$.
            \STATE Solve the unicast max-min power allocation problem in \eqref{unicast_max-min} to get the optimum unicast powers $\rho_{kg}^\star$.
            \STATE Initialize $\mathbf{P}_g$, $\forall g$ with the optimum unicast powers. \label{initialization}
            \FOR{$g = 1, \hdots, G$}
            \STATE Compute $\mathbf{R}_{\mathrm{int}, g} = \left(\sum\limits_{\substack{j = 1 \\ j \neq g}}^G\mathbf{H}_{j}\mathbf{H}_{j}^H + \frac{\sigma^2}{P_{\textrm{max}}}\mathbf{I}_{LN}\right)$.
            \STATE Compute $\bar{\mathbf{H}}_g = \mathbf{R}_{\mathrm{int}, g}^{-1}\mathbf{H}_g$.
            \STATE Set $\mathbf{R}_g \leftarrow \bar{\mathbf{H}}_g^H\left(\mathbf{P}_g\circ\widetilde{\mathbf{H}}_g\right)$.
            \FOR{$i = 1, \hdots, S$}
            \STATE Compute the desired signal gain $\textrm{DS}_{kg}$, $\forall k \in \{1, \hdots, K_G\}$ as the row sum of $\mathbf{R}_g$.
            \STATE Set $k_\mathrm{min} \leftarrow \argmin\limits_{k \in \{1, \hdots, K_G\}}\textrm{DS}_{kg}$. \label{k_min}
            \STATE Set $\textrm{UE}_\mathrm{min} \leftarrow \mathbf{R}_g^{\left(k_\mathrm{min}, k_\mathrm{min}\right)}$. \label{UE_min}
            \STATE Compute $\theta \leftarrow \angle \left(\textrm{DS}_{k_\mathrm{min}g} - \textrm{UE}_\mathrm{min}\right) - \angle \textrm{\,UE}_\mathrm{min}$. \label{theta}
            \STATE Update $\boldsymbol{\rho}_{\mathrm{rt},g}^{\left(k_\mathrm{min}\right)} \leftarrow re^{j\theta} \boldsymbol{\rho}_{\mathrm{rt},g}^{\left(k_\mathrm{min}\right)}$, and update $\mathbf{P}_g$. \label{rho_update}
            \STATE Update $\mathbf{R}_g \leftarrow \bar{\mathbf{H}}_g^H\left(\mathbf{P}_g\circ\widetilde{\mathbf{H}}_g\right)$.
            \ENDFOR
            \STATE Compute the desired precoder direction $\mathbf{d}_g$ as the row sum of $\mathbf{P}_g\circ\widetilde{\mathbf{H}}_g$.
            \STATE Compute $\bar{\mathbf{w}}_g = \mathbf{R}_{\mathrm{int}, g}^{-1}\mathbf{d}_g$.
            \STATE Set $\mathbf{w}_g \leftarrow \bar{\mathbf{w}}_g/\norm{\bar{\mathbf{w}}_g}$.
            \ENDFOR
            \STATE Scale the precoders $\mathbf{w}_g$, $\forall g$, such that the highest per-AP power satisfies the power constraint with equality.
\end{algorithmic}
\textbf{Output:} The multigroup multicast precoders $\mathbf{w}_g$, $\forall g$.
\end{algorithm}

\section{Numerical Evaluation} \label{results}

In this section, we use Monte Carlo simulations to evaluate the performance of the proposed optimization procedure and heuristic algorithm to solve the multigroup multicast MMF problem. We consider a cell-free network with $L$ APs deployed on a square grid in an area of $750\,\textrm{m} \times 750\,\textrm{m}$. A wrap-around topology is employed to mitigate boundary effects. Each AP is equipped with a half-wavelength-spaced uniform linear array of $N$ antennas. Unless otherwise indicated, we assume $G = 3$ multicast groups each comprising the same number of $K_g = K_G$ UEs, $\forall g$, that are randomly and uniformly distributed within the area of interest. We consider two cell-free network setups, the simulation parameters are summarized in Table \ref{params}. The channel between an arbitrary UE $k$ in group $g$ and AP $l$ is modelled by correlated Rayleigh fading as $\mathbf{h}_{kg,l} \sim \mathcal{N}_{\mathbb{C}}(\mathbf{0}, \mathbf{R}_{kg,l})$, where $\mathbf{R}_{kg,l} \in \mathbb{C}^{N \times N}$ represents the spatial correlation matrix, generated as in \cite{demir2021foundations}. The average channel gain, $\beta_{kg,l} = \frac{1}{N} \hspace{1pt}\textrm{tr}\hspace{-1pt}\left(\mathbf{R}_{kg,l}\right)$, is calculated using the 3GPP Urban Microcell model for generating large-scale fading coefficients with correlated shadowing among the UEs. The large-scale fading coefficients are thus given by
\begin{equation}
\beta_{kg,l} = -30.5 - 36.7 \textrm{log}_{10}\left(\frac{d_{kg,l}}{1\,\textrm{m}}\right) + F_{kg,l} \hspace{2pt}\textrm{dB},
\label{pathloss}
\end{equation}
where $d_{kg,l}$ denotes the distance between UE $k$ in group $g$ and AP $l$ and $F_{kg,l} \sim \mathcal{N}\left(0, 4^2\right)$ corresponds to the shadow fading. The shadowing is correlated between a single AP and different UEs as
\begin{equation}
\mathbb{E}\{F_{kg,l}F_{ij,n}\} = 
\begin{cases}
    4^22^{-\delta_{kgij}/9\,\textrm{m}} & \textrm{for } l = n, \\
    0 & \textrm{for } l \neq n,
\end{cases}
\label{shadowing}
\end{equation}
where $\delta_{kgij}$ is the distance between UE $k$ of group $g$ and UE $i$ of group $j$. Note that the correlation between the shadowing at different APs, which is the second case in \eqref{shadowing}, will be negligible because of the relatively larger inter-AP distances compared to the distances between the UEs.

\begin{table}
\vspace{0.03in}
\begin{center}
\caption{Network simulation parameters.}
\begin{tabular}{ |c|c| }
\hline
Area of interest & $750\,\textrm{m} \times 750\,\textrm{m}$ \\
Bandwidth & $20$\,MHz \\
Number of APs & $L = \{9, 4\}$ \\
Number of antennas per AP & $N = \{4, 8\}$ \\
Number of multicast groups & $G = 3$ \\
Number of UEs per group & $K_G = \{10, 30\}$ \\
Maximum AP transmit power & $P_{l,\mathrm{max}} = \{1, 2\}$\,W \\
Pathloss exponent & $\alpha = 3.67$ \\
DL noise power & $-94$\,dBm \\
\hline
\end{tabular}
\label{params}
\end{center}
\end{table}

\subsection{Performance Analysis} \label{perf_analysis}

In the following, the proposed successive elimination algorithm will be referred to as ``SEA'', while SDR followed by choosing the dominant eigenvector or Gaussian randomization as ``SDR-D'' and ``SDR-G'', respectively. The stopping criterion for both SDR-based algorithms and the DCA algorithm is chosen as $\epsilon = 0.1$. The elimination interval width for the SEA is set utilizing $\kappa = 0.96$. The penalty factor is adjusted to $\zeta = 30$.  The number of randomization vectors utilized for SDR-G is  $300$ and the dominant eigenvector solution of SDR-D is chosen as one of the candidate vectors. The number of iterations for phase alignment and UE emphasis scaling is set to $S = K$. We assume all groups have the same SINR target, that is $\eta_g = 1$, $\forall g$.

\begin{figure}
\setlength{\abovecaptionskip}{0.33cm plus 0pt minus 0pt}
\begin{subfigure}{.5\textwidth}
\centering
\includegraphics[scale=0.46]{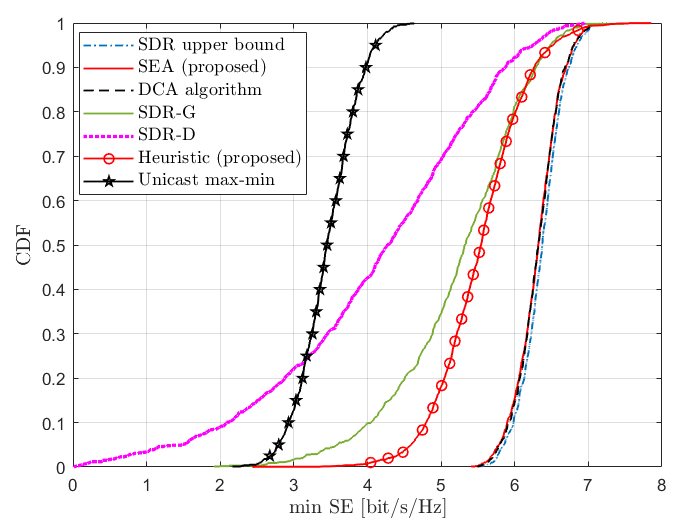}
\caption{$K = 30$ UEs.}
\label{k30}
\end{subfigure}
\begin{subfigure}{.5\textwidth}
\centering
\includegraphics[scale=0.46]{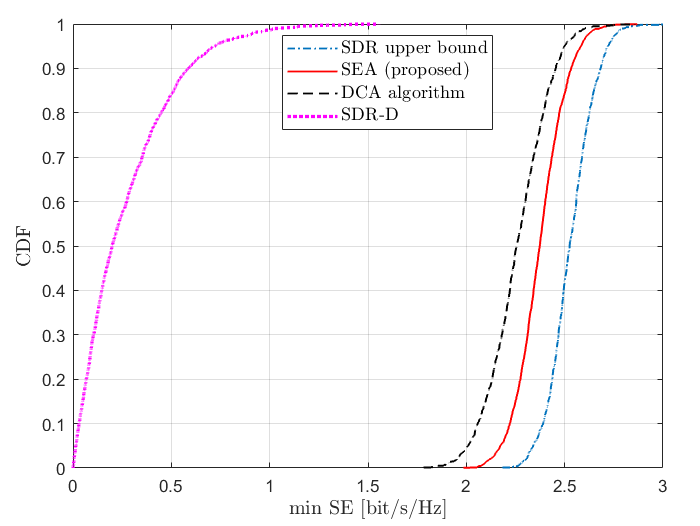}
\caption{$K = 90$ UEs.}
\label{k90}
\end{subfigure}
\caption{CDF of the minimum SE for $L = 9$, $N = 4$, and $P_{l,\mathrm{max}} = 1$\,W.}
\end{figure}

Figs.~\ref{k30} and \ref{k90} plot the cumulative distribution function (CDF) of the minimum SE for the proposed optimization procedure for $K = 30$ ($K_G = 10$) and $K=90$ ($K_G = 30$) UEs, respectively. The state-of-the-art DCA algorithm \cite{hsu2016joint} and previous SDR-based rank-1 solutions \cite{karipidis2008quality} as well as the SDR upper bound representing the higher-rank solution are shown for comparison. It is clear that the proposed SEA significantly outperforms previous rank-1 solutions based on SDR. For $K = 30$ UEs, both the SEA and DCA algorithms are able to achieve almost the same minimum SE as that of the SDR upper bound. For the dense case of $K = 90$ UEs, the proposed procedure achieves, on average, a $10.3$~bit/s/Hz increase in the sum-SE delivered to all UEs compared to the DCA algorithm, while requiring roughly the same computational time. This shows that the proposed procedure sets a new upper bound for what can be achieved with a rank-1 beamforming solution. Moreover, the figure shows that the gap between the proposed SEA and the DCA algorithm is larger in the lower-tail part of the CDF curve. In terms of the $90\,\%$-likely minimum SE, the proposed SEA achieves an increase of $7.9\,\%$ over the DCA algorithm. The reason is that in the worst-case situations, the gap between the local optimum achieved by the DCA algorithm and the global optimum can be large. On the other hand, the proposed procedure is seen to provide a more consistent and smaller gap to the global optimum. This can be inferred since the gap to the higher-rank upper bound is approximately constant.

The intuition behind the SE improvement is that the DCA algorithm follows a gradient descent concept that can only guarantee convergence to a stationary point of the original non-convex MMF problem. On the other hand, the proposed procedure starts with the optimum higher-rank solution that is delivered by SDR and then iteratively eliminates the higher-rank solutions by penalizing the direction of the eigenvector corresponding to the second-largest eigenvalue of these solution matrices. These eliminated solutions represent all the possible higher-rank solutions that can achieve an MMF SINR that is higher than or equal to that of the optimum rank-1 solution. In this way, the algorithm iterates until reaching the near-optimal rank-1 solution that lies in the orthogonal subspace to the directions of the penalized eigenvectors of all the higher-rank solutions of the previous iterations. The effect of limiting the search for the optimum rank-1 solution to the orthogonal subspace is small. The reason is that for each iteration, the solution matrices $\{\mathbf{W}_g\}_{g = 1}^G$ are Hermitian positive semidefinite matrices with distinct non-zero eigenvalues, which results in the eigenvectors of each matrix being orthogonal to each other. Consequently, the effect of the elimination procedure where we penalize the second strongest eigenvectors is minimal on subsequent iterations and thus produces a near-global optimum rank-1 solution to the NP-hard multigroup multicast MMF problem.

Further, we plot the CDF of the achievable minimum SE for the proposed heuristic algorithm for $K = 30$ UEs in Fig.~\ref{k30}. The optimum unicast max-min SE, which is used to initialize the heuristic algorithm, is shown as a baseline scheme for comparison. The results show that the phase alignment and UE emphasis scaling procedure offers substantial improvement compared to the baseline. In particular, a performance improvement in the median minimum SE is achieved for every UE from $3.45$~bit/s/Hz for the unicast scheme to $5.54$~bit/s/Hz for the multicast heuristic algorithm. Remarkably, the low-complexity heuristic is able to achieve significant gains over previous SDR-based optimization procedures that require extensive computations. Later, we will also show the immense improvement of the proposed heuristic in terms of computational requirements.

\begin{figure}
\centering
\setlength{\abovecaptionskip}{0.33cm plus 0pt minus 0pt}
\includegraphics[scale=0.46]{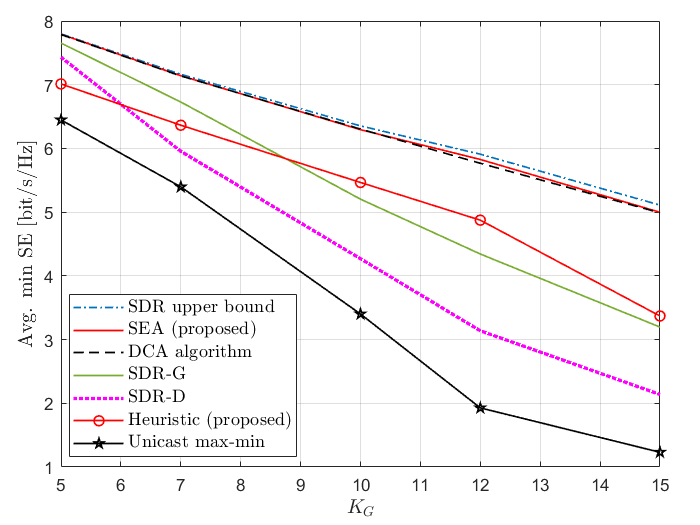}
\caption{Average minimum SE for different numbers of UEs, $L = 9$, $N = 4$, and $P_{l,\mathrm{max}} = 1$\,W.}
\label{avg_SE}
\end{figure}

\begin{figure}
\centering
\setlength{\abovecaptionskip}{0.33cm plus 0pt minus 0pt}
\includegraphics[scale=0.46]{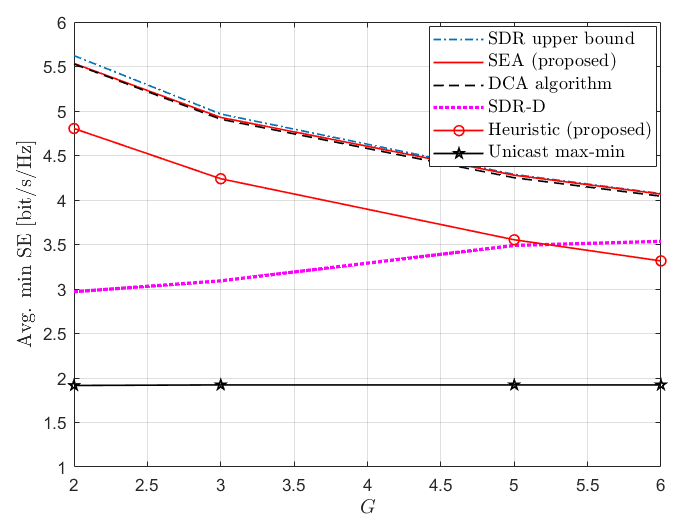}
\caption{Average minimum SE for different numbers of groups, $L = 4$, $N = 8$, $K = 30$ and $P_{l,\mathrm{max}} = 2$\,W.}
\label{avg_SE_against_G}
\end{figure}

Fig.~\ref{avg_SE} plots the average minimum SE versus the number of UEs per group $K_G$. It can be seen that the optimum unicast max-min solution presents a lower bound on the achievable average minimum SE via multicasting, which shows the possible improvements of multicast transmissions over unicasting when delivering the same data content to multiple UEs. The figure shows that the proposed SEA and the DCA algorithm maintain an average minimum SE that is close to the upper bound up to $K_G = 15$ UEs per group. The SDR-D and SDR-G provide relatively good performance for small $K_G$. The reason is that when the number of UEs is small, the solution matrices produced by SDR have ranks that are close to $1$. As a result, the approximations made by picking the dominant eigenvector or performing randomization do not sacrifice much to achieve an approximate rank-1 solution. As the number of UEs increase, a faster degradation in performance is seen for these approximation techniques compared to other algorithms. The proposed heuristic algorithm shows $5\,\%$, $28\,\%$, and $60\,\%$ improvement in the average minimum SE over SDR-G, SDR-D, and unicast max-min for $K_G = 10$ UEs per group, respectively.
Further, the heuristic algorithm is able to attain a constant gap to the optimum rank-1 solutions up to $K_G = 12$ UEs per group, which translates to about $87\,\%$ of the achievable average minimum SE of the optimum rank-1 solution for $K_G = 10$ UEs per group.
In Fig.~\ref{avg_SE_against_G}, we plot the average minimum SE against the number of groups $G$, for a fixed total number of $K = 30$ UEs. We utilize a different cell-free network setup to showcase the applicability of the proposed algorithms in different simulation settings. Similar behavior is seen for the proposed algorithms and benchmarks.
The gap between the proposed heuristic and the optimum rank-1 solutions is mainly due to the greedy nature of the proposed heuristic for phase alignment and UE emphasis scaling. That is, phase tuning and UE emphasis scaling is done successively rather than concurrently within each UE group. Such an approach brings in a significant reduction in computational requirements at the expense of a moderate reduction in the minimum SE.

\begin{figure}
\centering
\setlength{\abovecaptionskip}{0.33cm plus 0pt minus 0pt}
\includegraphics[scale=0.46]{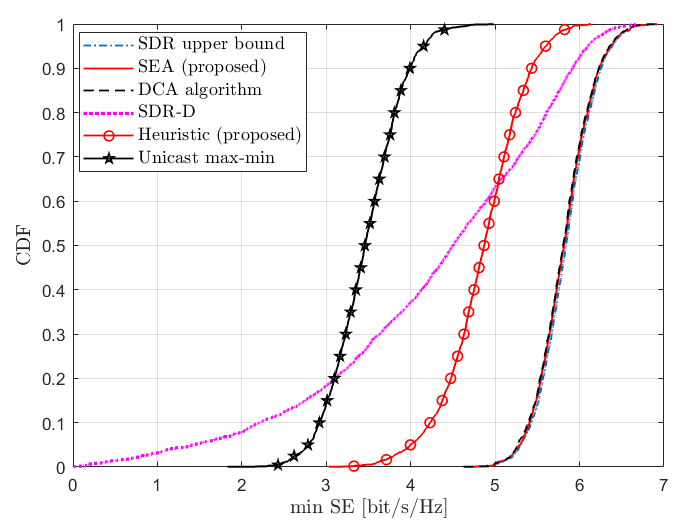}
\caption{CDF of the minimum SE with unequal group sizes, $L = 9$, $N = 4$, $K = 30$ and $P_{l,\mathrm{max}} = 1$\,W.}
\label{max_min_SE_unequal}
\end{figure}

Fig.~\ref{max_min_SE_unequal} plots the CDF of the minimum SE for $G = 6$ groups. To verify the applicability of the proposed algorithms to different settings, including a combination of co-located multicast and unicast transmissions, unequal group sizes are considered. Specifically, the numbers of UEs per group for the first four multicast groups are set to $K_1 = 10$, $K_2 = 8$, $K_3 = 7$ and $K_4 = 3$, whereas the last two groups have $1$ UE each, representing unicast transmissions. The results show that similar SE performance, to the case of equal-sized multicast groups, can be achieved for the proposed algorithms and benchmarks. It can be seen that the upper-tail part of the CDF curve becomes relatively better for SDR-D. The reason is that multicast groups of smaller sizes and unicast transmissions have a higher chance of obtaining a rank-1 solution directly after solving the relaxed SDP problem. When comparing the computational time of the proposed SEA to the DCA algorithm, a $54\,\%$ reduction in runtime is achieved in this scenario. More results on the computational complexity are provided in the next sections.

To evaluate the optimality gap to the global optimum, we perform a comparative analysis to the global \emph{Argument Cut Based Relaxation Branch-and-Bound} (ACR-BB) algorithm in \cite{lu2017efficient}. Due to the limited capability of the ACR-BB algorithm to simulate only single-cell single-group setups and its exponential complexity scaling, we perform the analysis considering a small simulation scenario with a single AP having $N = 6$ antennas deployed in an area of $250\,\textrm{m} \times 250\,\textrm{m}$. Table~\ref{ACRBB_comparison} compares the proposed SEA and the state-of-the-art DCA algorithm to the global optimum achieved by the ACR-BB algorithm for different numbers of UEs. The first two rows show the average and $90^{\textrm{th}}$ percentile of the difference in minimum SE in bit/s/Hz to the global optimum. It can be seen that around three times smaller difference to the global optimum is achievable with the proposed SEA. The following rows present the percentage of problem instances where the proposed SEA and the DCA algorithm are able to achieve a solution that is within a certain percentage of the global optimum. The proposed SEA can provide up to $25\,\%$ increase in the percentage of problem instances where a near-global optimal solution is achieved given a certain tolerance.

    \setcounter{table}{1}
    \begin{table}[h!]
    \begin{center}
    \caption{\centering Comparison to the global ACR-BB algorithm for $N = 6$ and $P_{\mathrm{max}} = 2$\,W.}
    \begin{tabular}{ |c|c|c|c|c|  }
    \hline
    \multirow{2}{*}{Performance measure} & \multicolumn{2}{c|}{SEA} & \multicolumn{2}{c|}{DCA} \\
    & $K = 20$ & $K = 30$ & $K = 20$ & $K = 30$ \\
    \hline
    Avg. diff. & $0.03$ & $0.07$ & $0.10$ & $0.20$\\
    $90\,\%$-likely diff. & $0.11$ & $0.22$ & $0.36$ & $0.61$\\
    Instances within $1\,\%$ & $81\,\%$ & $67\,\%$ & $68\,\%$ & $47\,\%$\\
    Instances within $2\,\%$ & $88\,\%$ & $76\,\%$ & $71\,\%$ & $58\,\%$\\
    Instances within $3\,\%$ & $94\,\%$ & $87\,\%$ & $77\,\%$ & $62\,\%$\\
    \hline
    \end{tabular}
    \label{ACRBB_comparison}%
    \end{center}
    \end{table}

\subsection{Convergence Behaviour}

\begin{figure}
\centering
\setlength{\abovecaptionskip}{0.33cm plus 0pt minus 0pt}
\includegraphics[scale=0.46]{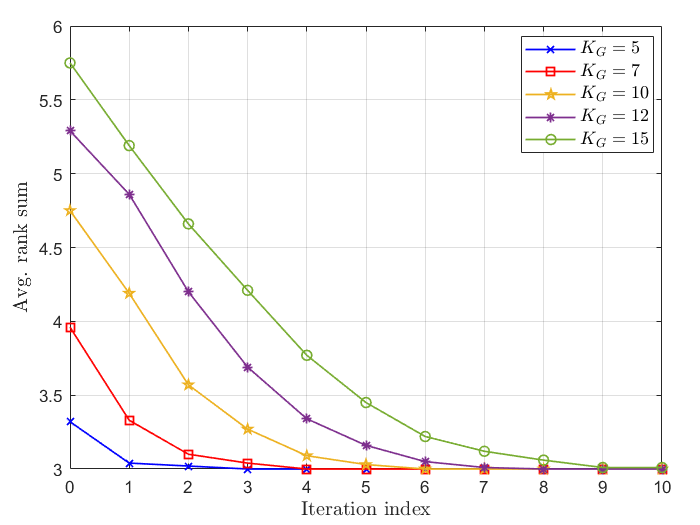}
\caption{Average sum of ranks for all group precoders versus the SEA iterations.}
\label{avg_rank_sum}
\end{figure}

In this subsection, we show results for the cell-free network with $L = 9$ APs and $N = 4$ antennas per AP. Similar behavior is obtained for other network settings. As previously stated, the proposed SEA works by eliminating the higher-rank solutions produced via SDR. Fig.~\ref{avg_rank_sum} plots the average sum of ranks of all group precoders against the iteration index of the elimination procedure for different numbers of UEs per group. The result shows a smooth reduction in the precoder ranks with each iteration on average. The curves saturate at a sum of ranks equal to $3$, which is the number of groups $G$, such that each group precoder has a rank of $1$. Note that the average sum of ranks is reduced by a value less than $1$ for every iteration. The reason is that in many situations, when eliminating a higher-rank solution, another solution of the same rank is produced via SDR, however, with a smaller minimum eigenvalue. Further, it is evident that as the number of UEs per group increases, the average ranks for the SDR solution matrices increase. This entails that more iterations are required to eliminate the higher-rank solutions produced via SDR.

\begin{figure}
\centering
\setlength{\abovecaptionskip}{0.33cm plus 0pt minus 0pt}
\includegraphics[scale=0.46]{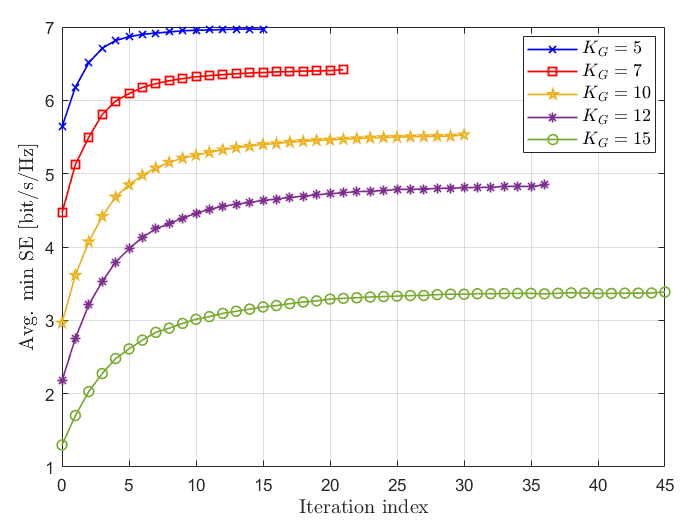}
\caption{Average minimum SE versus the phase alignment and UE emphasis scaling iterations.}
\label{avg_SE_VS_iterations_heuristic}
\end{figure}

Fig.~\ref{avg_SE_VS_iterations_heuristic} presents the average minimum SE achieved by the proposed heuristic algorithm versus the phase alignment and UE emphasis scaling iterations. The figure shows that there is a significant improvement in the average minimum SE, particularly in the first iterations. Afterwards, the achievable SE starts to saturate and no further improvement can be seen with increasing the number of iterations. This saturation point is seen to move right with increasing $K_G$, which implies that more iterations are needed to get the best possible minimum SE from the heuristic algorithm as the number of UEs increases. We would like to call attention to the fact that the main idea behind heuristic methods in the previous literature is to utilize a weighted sum of unicast precoders for each multicast group, and only optimize power allocation. We point out that the performance of these methods is similar to that of the initialization of our proposed heuristic, before performing the phase alignment and UE emphasis scaling procedure; that is, at iteration $0$. It is clear that a remarkable improvement is achieved with our novel low-complexity iterative procedure. We highlight that the performance improvement increases with increasing the number of UEs per group $K_G$. This verifies our earlier claim in Section \ref{sec_intro}, that previous heuristics perform relatively poorly when the multicast problem structure becomes more distant to that of the unicast problem.

For both Figs.~\ref{avg_rank_sum} and \ref{avg_SE_VS_iterations_heuristic}, the averaging is done over $100$ simulation instances with different UE locations and channel realizations.

\subsection{Complexity and Convergence Analysis} \label{complexity}

In this subsection, we evaluate the computational requirements of the proposed algorithms compared to the selected state-of-the-art benchmarks. The main computational burden for solving the MMF multigroup multicast beamforming problem with the proposed SEA lies in solving the corresponding relaxed QoS problem via SDP. A general-purpose solver that employs interior-point methods requires at most $\mathcal{O}(\sqrt{GLN})$ iterations to solve the problem, such that each iteration has a worst-case complexity of $\mathcal{O}(G^3(LN)^6 + (L + K)G(LN)^2)$ \cite{christopoulos2014weighted}. Note that actual runtimes grow far slower with $LN$ than this worst-case bound \cite{karipidis2008quality,luo2010semidefinite,christopoulos2014weighted}. In fact, a typical complexity of $\mathcal{O}(G^3(LN)^3)$ to $\mathcal{O}(G^3(LN)^4)$ per iteration is achievable depending on the structure of the problem at hand \cite{luo2010semidefinite,vandenberghe1996semidefinite,ye2011interior}, which is comparable to the complexity per iteration of the DCA algorithm discussed next. Further, a bisection search is done over the relaxed QoS problem to obtain the solution to the relaxed MMF problem. This bisection runs for $N_\mathrm{bis} = \mathrm{log}_2[(\gamma_\mathrm{up} - \gamma_\mathrm{lo})/\epsilon]$ iterations.\footnote{Given the SINR ranges at hand, we have observed that this translates to around $12$-$16$ iterations.} This procedure is repeated to eliminate the higher-rank solutions resulting from the SDP, with the difference that only a few bisection iterations\footnote{Considering the update of $\gamma_\mathrm{lo}$ on line~\ref{kappa_interval} in Algorithm~\ref{alg1}, typically, $5$-$6$ iterations are sufficient.} over the relaxed QoS problem are required since the upper bound can be chosen as the optimum objective achieved in the previous iteration as discussed in Section \ref{sec_SEA}.
On the other hand, solving a single convex subproblem of the DCA algorithm requires at most $\mathcal{O}(\sqrt{GLN + K})$ iterations, with a computational complexity of $\mathcal{O}((GLN + K)^3)$ per iteration \cite{tran2013conic,chen2017admm}. Starting from a feasible point, this procedure is then repeated until convergence to a stationary point of the original non-convex problem.\footnote{We have observed numerically that around $20$ iterations are needed to achieve the same $\epsilon$-accuracy as the proposed SEA.} As for the proposed heuristic, the main computational burden lies in calculation of the matrix inverse for $\mathbf{R}_{\mathrm{int},g} \in \mathbb{C}^{LN \times LN}$, $g = 1, \hdots, G$. The complexity for the rest of the procedure scales as $\mathcal{O}(KLN)$ for each multicast group. This results in an overall computational complexity of order $\mathcal{O}((G + 1)((LN)^3 + KLN))$.

\begin{figure}
\centering
\setlength{\abovecaptionskip}{0.33cm plus 0pt minus 0pt}
\includegraphics[scale=0.46]{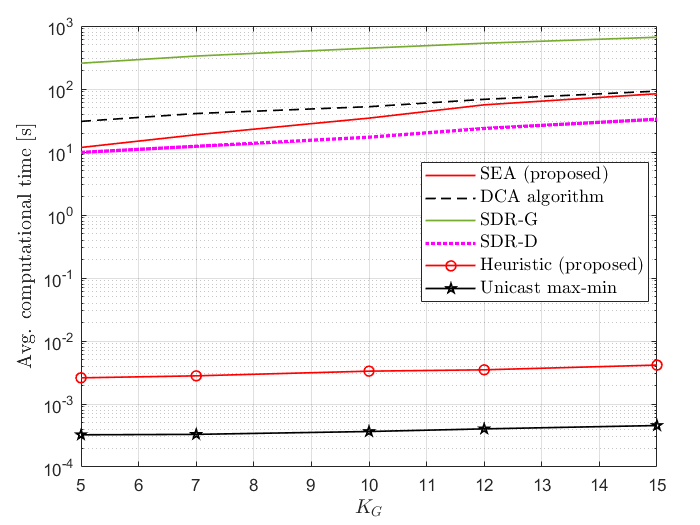}
\caption{Average computational time in seconds, $L = 9$, $N = 4$, and $P_{l,\mathrm{max}} = 1$\,W.}
\label{avg_runtimes}
\end{figure}

Since order expressions provide limited information about an algorithm's complexity in normal-sized scenarios due to the lack of scaling factors and lower-dimensional terms as well as rely on worst-case bounds which could lead to ill-informed conclusions, we supplement the analysis with numerical measurements for the average runtimes required by each algorithm. We use the same platform for performing all simulations for a fair comparison: a 4 core Intel(R) Core i5-10310U CPU with 1.7 GHz base frequency and 4.4 GHz turbo frequency. All programs are written in Matlab and utilize CVX \cite{cvx} to solve the optimization problems. Fig.~\ref{avg_runtimes} plots the average computational time against $K_G$. The figure shows that the proposed SEA achieves about $54\,\%$ and $35\,\%$ reduction in computational time compared to the state-of-the-art DCA algorithm for $K_G = 7$ and $K_G = 10$ UEs per group, respectively, and the reduction is particularly large when $K_G$ is small. This verifies the possibility of achieving the typical complexity per iteration for solving the associated SDPs. Moreover, the proposed heuristic algorithm is able to provide orders-of-magnitude reduction in computational time compared to the other multigroup optimization procedures, setting a record-low runtime for multicast beamforming design of $3.3\,$ms for $K_G = 10$ UEs per group. Hence, this algorithm could likely be directly used in real-time applications. In comparison to the unicast max-min scheme, the proposed heuristic strikes a balance in the inevitable trade-off between SE performance and computational complexity.

\begin{figure}
\centering
\setlength{\abovecaptionskip}{0.33cm plus 0pt minus 0pt}
\includegraphics[scale=0.46]{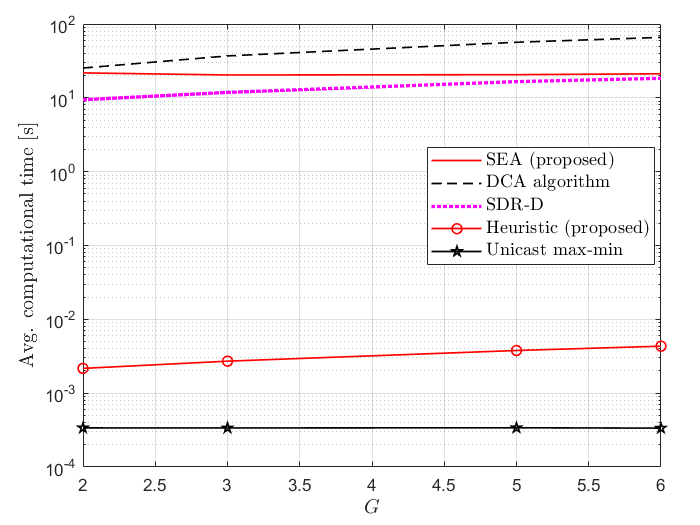}
\caption{Average computational time in seconds, $L = 4$, $N = 8$, $K = 30$, and $P_{l,\mathrm{max}} = 2$\,W.}
\label{avg_runtimes_against_G}
\end{figure}

In addition, Fig.~\ref{avg_runtimes_against_G} plots the average computational time against $G$. It can be seen that the proposed SEA maintains roughly a constant average computational time when the total number of UEs $K$ is fixed, regardless of the number of multicast groups. The reason is that although the complexity per iteration increases, the number of iterations required to eliminate the higher-rank solutions decreases along with the number of UEs per group $K_G$ (see Fig.~\ref{avg_rank_sum}). As for the DCA benchmark, a linear increase in the overall average computational time is observed with increasing the number of multicast groups. Starting from $G = 5$ multicast groups, the DCA algorithm requires more than three times the computational time of the proposed SEA. This shows that the proposed SEA is more capable of simultaneously serving an arbitrarily large number of multicast groups as compared to the benchmark, at no additional cost as long as the total number of UEs is fixed.

\section{Conclusions} \label{conc}

In this paper, the multigroup multicast MMF optimization problem has been revisited in the context of cell-free massive MIMO systems. We have proposed a novel iterative optimization procedure that is able to provide a near-global optimum rank-1 beamforming solution to this NP-hard problem. The proposed algorithm is compared against state-of-the-art SDR-based and SCA-based algorithms in multicast beamforming optimization. The numerical results show significant improvements in terms of SE performance and computational complexity while utilizing the same platform and software to solve optimization problems. The rationale behind the improvements is that the proposed procedure relies on eliminating high-rank solutions while imposing a negligible effect on the optimal achievable SE, whereas previous methods rely on approximations that degrade the performance, particularly when the number of UEs grows large. The significance of the proposed algorithm is that it can be adopted for any type of problem where SDR is utilized and a low-rank solution is desirable, which is a wide class of optimization problems. Moreover, we have developed a low-complexity phase alignment and UE emphasis scaling heuristic algorithm that is able to achieve about $87\,\%$ of the minimum SE of the optimum rank-1 solution with orders-of-magnitude lower computational time. The heuristic algorithm enables a real-time implementation of multigroup multicast beamforming. For future work, the impact of imperfect CSI and problem-specific optimization procedures (rather than general-purpose solvers) can be considered for performance evaluation.

\section*{References}
\renewcommand{\refname}{ \vspace{-\baselineskip}\vspace{-1.1mm} }
\bibliographystyle{ieeetr}
\bibliography{papercites}

\end{document}